\documentclass[prd,aip,graphicx,nofootinbib]{revtex4-1}

\usepackage{natbib}
\usepackage{bm}
\usepackage{amscd, amsfonts, amssymb}
\usepackage{epsfig}
\usepackage[mathscr]{eucal}
\usepackage{verbatim}
\usepackage{latexsym}
\usepackage{enumerate}

\draft 

\begin{document}





\title{Elko Field Theory and the Nature of Elko Darkness} 








\author{Adam Gillard}

\email[]{adam.gillard@pg.canterbury.ac.nz}




\affiliation
{Department of Physics and Astronomy,
University of Canterbury,
Private Bag 4800,
Christchurch 8140,
New Zealand}





\date{\today}

\begin{abstract}
We show that the Elko Lagrangian dark matter candidate is gauge-invariant under local gauge transformations and that non-abelian gauge invariance can be set up quite naturally. This leads naturally to Elko symmetry currents. These Elko symmetry currents can then be coupled with the symmetry currents of the Electroweak theory to form interaction Hamiltonian densities, thus providing a natural mechanism for Elko particles to interact directly with Standard Model matter in addition to the Higgs particle. This could have profound implications for the detection of Elko at the LHC. We also show that in a certain sense Elko fields are non-local even along the  \textit{axis of locality}. This may make direct detection of Elko difficult. We propose that Elko would therefore still seem dark to us, as far as direct detection efforts are concerned. We propose however, that the allowed gauge interactions between Elko and the electroweak sector of the Standard Model may result in the possibility of new experiments to be devised to look for Elko \textit{indirectly} via studying the Standard Model particles that are involved in an Elko interaction. Such new experiments could complement existing ideas of developing experiments to detect the Elko particles based on their interaction with the Higgs particle.
\end{abstract}

\pacs{}

\keywords{Dark Matter, Elko Fields, Non-Local}

\maketitle 







%
%
%
%
%
\section{Introduction}
Ever since the discovery of evidence for dark matter by Fritz Zwicky in 1933 [\onlinecite{zwicky1933rotverschiebung}], people have been proposing different types of particles as candidates to account for this dark matter [\onlinecite{hinshaw2009five,peccei1977constraints,weinberg1978new,maiani1986effects,zavattini2006experimental,fukuda1998measurements,akerib2004installation,foot2010relevance,mambrini2010kinetic,archambault2009dark}]. 
In 2005 a new spin-1/2 massive non-local mass dimension one fermionic quantum field was introduced called \textit{Eigenspinoren des Ladungskonjugationsoperators} commonly referred to simply as \textit{Elko} [\onlinecite{ahluwalia2005spin}]. R. da Rocha and W.A. Rodrigues Jr showed that Elko spinors belong to a class of flagpole spinors which appear in Lounesto's classification of spinors [\onlinecite{da2006elko}][\onlinecite{lounesto2001clifford}]. Elko spinors have since been the subject of increasing interest and research. In 2006, C.G. Bohmer proposed that Elko spinors could act as sources of curvature and torsion [\onlinecite{boehmer2007einstein2}]. Boehmer also put forward the idea that Elko spinors help to solve the general problem concerning how Maxwell fields can be connected to Einstein-Cartan theory with minimal coupling. In the following year, da Rocha and J.M. Hoff da Silva constructed an algebraic way of relating Dirac spinors to Elko spinors [\onlinecite{da2007dirac2}]. In 2009 da Rocha and Hoff da Silva also went on to show that this algebraic mapping between Elko and Dirac spinors could be used in the process of deriving the Quadratic Spinor Lagrangian from the Einstein-Hilbert, Einstein-Palatini and Holst actions [\onlinecite{da2009elko}]. Elko spinors have also been proposed as being involved in the driving of inflation [\onlinecite{boehmer2007einstein2}][\onlinecite{shankaranarayanan2009if}]. Yet another use for the Elko spinors was introduced in 2010 by H. Wei. He gave a method of reconstructing spinor dark energy from cosmological observations when using Elko spinors [\onlinecite{wei2010couple}]. In 2010, G. Chee looked at Elko spinors in the context of a de Sitter universe [\onlinecite{chee2010stability}]. Also, L. Fabbri examined Elko spinor field interactions through contorsion with their own spin density. He also examined the most general dynamical theory for Elko spinors and examined some implications for torsional $f(R)$ theories. Fabbri has also looked into the problem of causal propagation for spinor fields and showed that the problem of causal propagation for Elko spinor fields is always solvable. Fabbri has also considered the role of Elko spinors in conformal gravity [\onlinecite{fabbrimost,fabbri2010causal,fabbri2010zero,fabbri2010causality,fabbri2010most,fabbri2011conformal}].

Much attention has also been given to the corresponding Elko \textit{quantum} fields, including a recent analysis by M. Dias, F. de Campos and da Silva on the prospects of detecting Elko particles at the LHC via interactions with the Higgs boson [\onlinecite{Dias:arXiv1012.4642}].

The original version of an Elko field was presented in 2005 and was non-local [\onlinecite{ahluwalia2005spin}], but in 2008 a local version was presented [\onlinecite{ahluwalia2008local}]. Understanding the underlying symmetry principles behind the Elko quantum field has proved to be a difficult task. It was discovered that Elko fields do not respect all of the symmetries of the Lorentz group [\onlinecite{Conferenceproceedings,ahluwalia2010elko,gillardmartinpaperv2}]. In [\onlinecite{ahluwalia2010elko}] it was proposed that Elko fields, although violating rotational symmetries, have an axis of locality. It has been pointed out by C.Y. Lee [\onlinecite{lee2010elko}] that, when restricted to 1+1 dimensions, Elko fields are local.

Since their introduction in 2005, Elko fields have always been put forward as a dark matter candidate. The discovery of an axis of locality is also consistent with some ideas on the properties of dark matter [\onlinecite{land2007axis,samal2009signals,frommert2010axis}].

 It is the darkness of Elko fields that the present paper is mainly about. Specifically, the central concept that we wish to communicate to the reader in this paper, is that the Elko Lagrangian is gauge-invariant. Elko doublets can be constructed that are invariant under local $SU(2)\times U(1)$ gauge transformations, which give rise to symmetry currents that can then be coupled to the symmetry currents of the electroweak sector of the Standard Model. 

The direct implication of this is that Elko fields are not dark with respect to Standard Model matter on the basis of not fitting into the Standard Model doublets because of having mass dimension one in contrast to Dirac fields which are of mass dimension three halves\footnote{At the spinor level Elko spinors, being eigenspinors of the finite-dimensional charge conjugation operator, show darkness with respect to the usual $U(1)$ gauge transformations but here in this paper we are considering Elko at the \textit{quantum field} level, where the elements of spinors are regarded as simply coefficient functions accompanying the creation and annihilation operators, and so do not enter the picture when considering gauge invariance of an Elko \textit{quantum field} Lagrangian.}. We suspect that if Elko fields \textit{do} behave as though they are dark, it is probably due to the non-local aspects of Elko. If the attribute of Elko fields having a preferred direction leads to difficulties in experimentally determining quantities such as the position and mass of Elko particles, then Elko would remain elusive and seemingly dark to us; at least as far as direct detection is concerned. Furthermore, as we show in this paper, Elko fields have an element of non-locality to them even along the axis of locality. In order to devise experiments to detect Elko indirectly by studying the Standard Model particles that are involved in an Elko interaction, a further study of Elko gauge interactions with Standard Model matter is needed. For now we will restrict our attention to the stated objectives of this paper: to show Elko local gauge invariance, leading to the conclusion that Elko should be able to interact with Standard Model particles via the usual electroweak gauge quanta of the Standard Model, and that Elko fields have an element of non-locality even along the axis of locality. We again emphasize that in saying this, we are \textit{not} saying that Elko ceases to be viable in terms of Elko being a dark matter candidate. The property of Elko having an axis of locality together with our observation that Elko carries an additional element of non-locality could be the defining properties that makes Elko fields appear dark, as far as direct detection efforts are concerned. 

In this paper, we start with a review of Elko fields and their interactions as viewed in [\onlinecite{ahluwalia2005spin}][\onlinecite{ahluwalia2008local}][\onlinecite{ahluwalia2010elko}][\onlinecite{ahluwalia2010very}] and include a review of the darkness aspects of Elko that were presented in the cited papers, along with the presented allowed interactions given the claims that Elko particles are dark with respect to all Standard Model particles apart from the Higgs boson. After the Elko field review, we show how the Elko Lagrangian can be invariant under gauge transformations, and the Elko fields therefore admit interactions with Standard Model gauge quanta. We then discuss how and why these results contradict results in the existing literature, which essentially comes down to the incompleteness of the Elko Field Theory. 

Next, we show that in a certain sense the Elko field is non-local even along the axis of locality. We propose that this provides the most likely grounds for continuing to argue that Elko may be very difficult to see, and apparently be dark, even though Elko particles should be able to interact with Standard Model gauge quanta.

We finish the paper by discussing the implications of our research for Elko Fields and indicate possible directions that future research in this area of theoretical physics could take in view of the results presented here.
\section{Elko Field Review and Elko Darkness} \label{sec:elkoreviewanddarkness}
We here include a short review of Elko fields to make this paper more self-contained and easily accessible to a wider audience. The term \textit{Elko field} has now come to take on more than one meaning. We will comment on both. The first meaning of the term \textit{Elko field} refers to fields $\Lambda(x)$ [\onlinecite[p.4]{ahluwalia2008local}]:
\begin{equation}
\Lambda(x)=\int\frac{d^3\mathbf{p}}{(2\pi)^3\sqrt{2mE(\mathbf{p})}}\sum_{h}[e^{-ip\cdot x}\xi(\mathbf{p},h)a(\mathbf{p},h)+e^{ip\cdot x}\zeta(\mathbf{p},h)b^{\dagger}(\mathbf{p},h)]
\end{equation}
where $a(\mathbf{p},h)$ are anti-Elko annihilation operators, $b^{\dagger}(\mathbf{p},h)$ are Elko creation operators, $h$ is a two-valued discrete index and $\xi(\mathbf{p},h)$ and $\zeta(\mathbf{p},h)$ are 4-dimensional spinors that transform according to the $(\frac{1}{2},0)\oplus(0,\frac{1}{2})$ class of representations of the Lorentz group. The rest spinors $\xi(\mathbf{0},h)$ and $\zeta(\mathbf{0},h)$ are defined  by having the form
\begin{equation}
\label{eqn:elkorestspinorgeneralform}
\left(\begin{array}{cccc}
\eta\Theta\phi(\mathbf{0})\\
\phi(\mathbf{0})
\end{array}\right)
\end{equation}
where $\phi(\mathbf{0})$ are eigenspinors of the helicity operator $\frac{1}{2}\bm{\sigma}\cdot \hat{\mathbf{p}}$. The phase $\eta$ is taken to be imaginary. The boosted spinors are obtained from the rest spinors by application of the boost matrix. The $\xi(\mathbf{p},h)$ are also taken to be eigenspinors of the $(\frac{1}{2},0)\oplus(0,\frac{1}{2})$ charge conjugation operator with eigenvalue $+1$ and the $\zeta(\mathbf{p},h)$ are eigenspinors of the same charge conjugation operator with eigenvalue $-1$. Elko spinors have a distinct dual from that of Dirac spinors. Elko spinors have duals 
$\stackrel{\neg}\xi(\mathbf{p},h)$ and $\stackrel{\neg}\zeta(\mathbf{p},h)$ defined by
\begin{equation}
\stackrel{\neg}\xi(\mathbf{p},\pm)\equiv \mp i\xi^{\dagger}(\mathbf{p},\mp)\gamma^0\quad\textrm{and}\quad \stackrel{\neg}\zeta(\mathbf{p},\pm)\equiv\mp i\zeta^{\dagger}(\mathbf{p},\mp)\gamma^0
\end{equation}
respectively. The Elko spinors together with the Elko dual spinors satisfy the usual orthonormality and completeness relations. Their spin sums, in contrast, produce something new. They are [\onlinecite[p.3]{ahluwalia2010elko}]:
\begin{equation}
\sum_{h}\xi(\mathbf{p},h)\stackrel{\neg}\xi(\mathbf{p},h)=m[\gamma^5\gamma_{\mu}g^{\mu}+\mathbb{I}]
\end{equation}
\begin{equation}
\sum_h\zeta(\mathbf{p},h)\stackrel{\neg}\zeta(\mathbf{p},h)=m[\gamma^5\gamma_{\mu}g^{\mu}-\mathbb{I}]
\end{equation}
where $g^{\mu}=(0,\mathbf{g})$ and 
\begin{equation}
\mathbf{g}=\frac{-1}{\sin(\theta)}\frac{\partial\hat{p}}{\partial\phi}=(\sin(\phi),-\cos(\phi),0).
\end{equation}
The appearance of the spacelike four-vector $g^{\mu}$ is what is responsible for Elko having a preferred axis directly giving rise to the notion of an axis of locality. The $\theta$ and $\phi$ are angles that parametrise $\hat{\mathbf{p}}$. The $\gamma_{\mu}$ and $\gamma^5$ are constant $4\times4$ matrices, which in the chiral representation take the specific form
\begin{eqnarray}
\gamma_0=\left(\begin{array}{cccc}
0&1\\
1&0
\end{array}\right),\quad \bm{\gamma}=\left(\begin{array}{cccc}
0&\bm{\sigma}\\
-\bm{\sigma}&0
\end{array}\right),\quad \gamma^5=\left(\begin{array}{cccc}
1&0\\
0&-1
\end{array}\right)
\end{eqnarray}
where each entry above represents a $2\times2$ sub-matrix in the obvious way, and where the $\sigma_i$ are the standard Pauli matrices
\begin{eqnarray}
\sigma_1=\left(\begin{array}{cccc}
0&1\\
1&0
\end{array}\right),\quad \sigma_2=\left(\begin{array}{cccc}
0&-i\\
i&0
\end{array}\right), \quad \sigma_3=\left(\begin{array}{cccc}
1&0\\
0&-1
\end{array}\right).
\end{eqnarray}

The Elko field $\Lambda(x)$ has an associated dual field $\stackrel{\neg}\Lambda(x)$, which takes the form [\onlinecite[p.4]{ahluwalia2008local}]
\begin{equation}
\stackrel{\neg}\Lambda(x)=\int\frac{d^3\mathbf{p}}{(2\pi)^3\sqrt{2mE(\mathbf{p})}}\sum_h[e^{ip\cdot x}\stackrel{\neg}\xi(\mathbf{p},h)a^{\dagger}(\mathbf{p},h)+e^{ip\cdot x}\stackrel{\neg}\zeta(\mathbf{p},h)b(\mathbf{p},h)].
\end{equation}
This dual is fundamentally different from the Dirac dual field $\bar{\psi}(x)=\psi^{\dagger}(x)\gamma_0$ because the Dirac dual is based on the adjoint of the Dirac field, which itself is also a well-defined quantum field operator. The dual Elko field $\stackrel{\neg}\Lambda(x)$ however, is based on the dual spinors which have opposite $h$-values to that of their accompanying creation and annihilation operators. Thus the physical interpretation of $\stackrel{\neg}\Lambda(x)$ is more obscure.

The Lagrangian density for free Elko fields takes the Klein-Gordon form [\onlinecite[p.5]{ahluwalia2008local}]:
\begin{equation}
\mathcal{L}^\Lambda(x)=\partial_{\mu}\stackrel{\neg}\Lambda\partial^{\mu}\Lambda-m^2\stackrel{\neg}\Lambda\Lambda.
\end{equation}
By inspection we see that the Elko field has mass dimension one, in contrast to the Dirac fields which have mass dimension three halves. The Elko conjugate momentum $\Pi(x)$ is defined by 
\begin{equation}
\Pi(x)\equiv\frac{\partial\mathcal{L}^\Lambda}{\partial\dot{\Lambda}}=\frac{\partial}{\partial t}\stackrel{\neg}\Lambda(x).
\end{equation}
For more details, we refer the reader to the literature [\onlinecite{ahluwalia2005spin}][\onlinecite{ahluwalia2008local}][\onlinecite{ahluwalia2010elko}]. Here however, we point out that for the Elko field $\Lambda(x)$ the anticommutators hold in the preferred direction: 
\begin{equation}
\{\Lambda(\mathbf{x},t),\Pi(\mathbf{x}',t)\}=i\delta^3(\mathbf{x}-\mathbf{x}')\mathbb{I}
\end{equation}
\begin{equation}
\{\Lambda(\mathbf{x},t),\Lambda(\mathbf{x}',t)\}=\mathbb{O}\quad\textrm{and}\quad \{\Pi(\mathbf{x},t),\Pi(\mathbf{x}')\}=\mathbb{O}.
\end{equation}
The more recent Elko-type field operator to be presented, may be found in [\onlinecite{ahluwalia2010very}]. This field $\Upsilon(x)$ is different from the Elko field $\Lambda(x)$. The rest spinors $\rho(\mathbf{0},h)$ and $\varrho(\mathbf{0},h)$ are numerically identical to the Elko rest spinors which have the general form given by Eqn.\ (\ref{eqn:elkorestspinorgeneralform}) but in contrast to the boosted Elko spinors, the boosted spinors $\rho(\mathbf{p},h)$ and $\varrho(\mathbf{p},h)$ are obtained from the rest spinors  by multiplying by the matrix
\begin{equation}
\left(\begin{array}{cccc}
\sqrt{\frac{m}{E-p_z}}&\frac{p_x-ip_y}{\sqrt{m(E-p_z)}}&0&0\\
0&\sqrt{\frac{E-p_z}{m}}&0&0\\
0&0&\sqrt{\frac{E-p_z}{m}}&0\\
0&0&-\frac{p_x+ip_y}{\sqrt{m(E-p_z)}}&\sqrt{\frac{m}{E-p_z}}
\end{array}\right).
\end{equation}
The spinors $\rho(\mathbf{p},h)$ and $\varrho(\mathbf{p},h)$ transform under the group generated by $SIM(2)$ adjoined with the four spacetime translations. The Lie algebra associated with $SIM(2)$ is given by
\begin{equation}
[T_1,T_2]=0,\quad [T_1,K_z]=iT_1,\quad [T_2,K_z]=iT_2
\end{equation}
\begin{equation}
[T_1,J_z]=-iT_2,\quad [T_2,J_z]=iT_1,\quad [J_z,K_z]=0
\end{equation}
where the generators $T_1$ and $T_2$ are defined by 
\begin{equation}
T_1\equiv K_x+J_y,\quad\textrm{and}\quad T_2\equiv K_y-J_x.
\end{equation}
with the $J_i$ and $K_i$ being the usual generators of rotations and boosts respectively for the Lorentz group. The associated Elko-type field $\Upsilon(x)$ is then:
\begin{equation}
\Upsilon(x)=\int\frac{d^3\mathbf{p}}{(2\pi)^3\sqrt{2mE(\mathbf{p})}}\sum_h[e^{-ip\cdot x}\rho(\mathbf{p},h)a(\mathbf{p},h)+e^{ip\cdot x}\varrho(\mathbf{p},h)b^{\dagger}(\mathbf{p},h)].
\end{equation}
The dual spinors and dual field operator $\stackrel{\neg}\Upsilon(x)$ are all defined analogously to the Elko field $\Lambda(x)$. $\Upsilon(x)$ also has an axis of locality for similar reasons as the $\Lambda(x)$ field. The Lagrangian density $\mathcal{L}^{\Upsilon}(x)$, for $\Upsilon(x)$, is also Klein-Gordon in form:
\begin{equation}
\mathcal{L}^{\Upsilon}(x)=\partial_{\mu}\stackrel{\neg}\Upsilon\partial^{\mu}\Upsilon-m^2\stackrel{\neg}\Upsilon\Upsilon,
\end{equation}
giving the fermionic field $\Upsilon(x)$ Elko's characteristic mass dimension one, making it very different from the normal Dirac-type fermionic fields. The operators $b^{\dagger}(\mathbf{p},h)$, for both the $\Lambda(x)$ field and the $\Upsilon(x)$ field can be taken as either distinct from the operator $a^{\dagger}(\mathbf{p},h)$, or the same as the operator $a^{\dagger}(\mathbf{p},h)$. So in principle some Elko particles may have distinct antiparticles while other Elko particles may be their own antiparticles.

Elko quantum fields were introduced in [\onlinecite{ahluwalia2005spin}] as a prime candidate for dark matter, having no Standard Model interactions other than with the Higgs particle. The associated interaction Lagrangian was given in [\onlinecite[p.44]{ahluwalia2005spin}] for an Elko field $\eta(x)$ and a Higgs doublet $\phi(x)$ as:
\begin{equation}
\label{eqn:elkohiggsinteractionaa}
\mathcal{L}_{\phi\eta}^{\textrm{int}}(x)=\lambda_{\textrm{E}}\phi^{\dagger}(x)\phi(x)\stackrel{\neg}\eta(x)\eta(x)
\end{equation}
with more interactions of this form possible if more scalar fields exist in nature. The symbol $\lambda_{\textrm{E}}$ is a dimensionless coupling constant. An Elko quartic self interaction was also introduced [\onlinecite[p.44]{ahluwalia2005spin}] of the form
\begin{equation}
\mathcal{L}^{\textrm{self}}(x)=\alpha_{\textrm{E}}[\stackrel{\neg}\eta(x)\eta(x)]^2
\end{equation}
with $\alpha_{\textrm{E}}$ another dimensionless coupling constant. In [\onlinecite[p.44]{ahluwalia2005spin}] Ahluwalia and Grumiller also state that the Elko field could couple to an abelian gauge field, via the field strength tensor $F_{\mu\nu}$ associated to the gauge field. Such an interaction has the form
\begin{equation}
\mathcal{L}_{\eta F}^{\textrm{int}}=\epsilon_{\textrm{E}}\stackrel{\neg}\eta(x)[\gamma^{\mu},\gamma^{\nu}]\eta(x)F_{\mu\nu}(x).
\end{equation}
Ahluwalia and Grumiller argued however, that the coupling constant $\epsilon_{\textrm{E}}$ would have to be vanishingly small because terms like these generate an effective mass for the photon but the possible mass of a photon has been experimentally severely constrained. Ahluwalia and Grumiller did not put the Elko field through the gauge process, but state [\onlinecite[p.44]{ahluwalia2005spin}] that \textit{the Elko field is neutral with respect to local $U(1)$ gauge transformations.} They then conclude that the Elko-Higgs interaction of Eqn.\ (\ref{eqn:elkohiggsinteractionaa}) is the dominant interaction between Elko particles and the Standard Model particles.

In [\onlinecite[p.61]{ahluwalia2005spin}], Ahluwalia and Grumiller clarify that the darkness of Elko fields is due to the fact that the mass dimensionality of the Elko field is one and not three halves which ``forbids a large class of interactions with gauge and matter fields of the Standard Model while allowing for an interaction with the Higgs field.'' 

In [\onlinecite[p.7]{ahluwalia2008local}], a stronger statement is made concerning Standard Model gauge transformations:
\\
``$\mathcal{L}^{\Lambda}(x)$ and $\mathcal{L}^{\lambda}(x)$ do not carry invariance under Standard Model gauge transformations.''

On the same page however, it was also stated that the Elko fields, although having Lagrangians that are not invariant under gauge transformations, are not forced to be self-referentially dark. The Elko spinors could undergo abelian gauge transformations of the form
\begin{equation}
\chi(\mathbf{p})\rightarrow \textrm{exp}[iM\alpha(x)]\chi(\mathbf{p})
\end{equation}
if and only if $M$ is the $4\times4$ matrix $\gamma_0$. Any non-abelian generalizations would have to retain this $\gamma_0$.

The reason for claiming that the Elko \textit{spinor} Lagrangian is not $U(1)$ gauge invariant, is that we have $S(C)\chi=\pm \chi$ but $S(C)\chi'\neq\pm\chi'$ where $\chi'=e^{i\alpha}\chi$ for some $\alpha$. Here $S(C)$ is the charge conjugation operator belonging to the chiral representation. Ahluwalia et.\ al.\ present $S(C)$ as being antilinear [\onlinecite[p.10]{ahluwalia2005spin}][\onlinecite[p.2]{ahluwalia2008local}][\onlinecite[p.5]{ahluwalia2010very}].

We will discuss these points after we demonstrate a counter claim, namely, that the Elko Lagrangians \textit{can} admit Standard Model gauge transformations. 

We have reviewed some of the main defining results for Elko fields. We now move on to show that the Elko Lagrangian, when expressed in terms of Elko field operators, rather than just spinors, can be made gauge invariant by an appropriate choices of covariant derivative coupled with the most natural way of filling a hole in the Elko Field Theory, the theory of which we claim is incomplete.
\section{Elko $U(1)$ Gauge Transformation}
In this section, we examine under what conditions $U(1)$ gauge invariance is possible for the Elko Lagrangian. The same things could also be said of the cousin field $\Upsilon(x)$. The usual Elko Lagrangian for free particles takes the form:
\begin{equation}
\mathcal{L}=\partial_{\mu}\stackrel{\neg}\Lambda\partial^{\mu}\Lambda-m^2\stackrel{\neg}\Lambda\Lambda.
\end{equation}
If we replace the partial derivatives by covariant derivatives $D^{\mu}$, of the usual form as used in Quantum Electrodynamics;
\begin{equation}
D^{\mu}=\partial^{\mu}+iqA^{\mu}
\end{equation}
for some vector field $A^{\mu}$, and further, define:
\begin{equation}
D^{\mu'}=\partial^{\mu}+iqA^{\mu}+iq\partial^{\mu}\chi
\end{equation}
for some scalar field $\chi$ and
\begin{equation}
\Lambda'=e^{-iq\chi}\Lambda,
\end{equation}
it then follows that 
\begin{equation}
D^{\mu'}\Lambda'=e^{-iq\chi}D^{\mu}\Lambda.
\end{equation}
We feel that it is worth emphasizing again at this point that since we are not looking at a Lagrangian composed of spinors, but, in contrast, are looking at an Elko Lagrangian operator acting on the Hilbert space composed of \textit{quantum fields}, the Elko spinor coefficient functions do not come directly into the analysis presented here, so do not have a bearing on the gauge invariance or non-gauge invariance of an Elko \textit{quantum field} Lagrangian. 

In order to be able to say whether the Elko Lagrangian is gauge invariant, we need to know what the effect of the Elko dual operation is on the  product of a non-Elko operator $A$ with an Elko operator $\Lambda$. The Elko Field Theory as it currently stands does not say anything about how the Elko dual operation should affect the product of Elko opeartors with non-Elko operators. This is an important aspect in which the Elko Field Theory is incomplete, leaving it an open question as to how to best plug the holes in the theory. We propose plugging this hole in Elko Field Theory by defining
\begin{equation}
\stackrel{\neg}{A\Lambda}\equiv\stackrel{\neg}\Lambda A^{\dagger}.
\end{equation}
We believe the way we have defined the operation is the most natural way of plugging this hole in the Elko Field Theory. If we hold to this, then the Elko Lagrangian $\mathcal{L}(\Lambda,\stackrel{\neg}\Lambda,D_{\mu}\Lambda,D_{\mu}\stackrel{\neg}\Lambda)$ is $U(1)$ gauge invariant, since we now have
\begin{equation}
\label{eqn:criticalforelkogaugeinvariance}
\stackrel{\neg}\Lambda'=\stackrel{\neg}\Lambda e^{iq\chi}
\end{equation}
and 
\begin{equation}
\stackrel{\neg}{D_{\mu}'\Lambda'}=\stackrel{\neg}{D_{\mu}\Lambda}e^{iq\chi}.
\end{equation}
This observation is new, and the Elko dual of such products were not discussed in the original papers. In the next section we take multiplets of Elko fields and examine the general non-abelian gauge symmetries. We will show that Elko Lagrangians can be invariant under non-abelian Standard Model gauge transformations like those in Quantum Chromodynamics and the Electroweak Theory.
\section{Elko Non-Abelian Gauge Transformations}
In this section, we consider Lagrangian's with Elko multiplets $\bm{\Lambda}^T=(\Lambda_1,\Lambda_1,\cdots,\Lambda_N)$, transforming under the standard $N$-dimensional representation of the group $SU(N)$. The free particle Elko Lagrangian then takes the form
\begin{equation}
\label{eqn:fourpointoneinelkosolopaper}
\mathcal{L}=\partial_{\mu}\bm{\stackrel{\neg}\Lambda}\partial^{\mu}\bm{\Lambda}-m^2\bm{\stackrel{\neg}\Lambda}\bm{\Lambda}.
\end{equation}
Consider an $SU(N)$ gauge transformation of the form
\begin{equation}
\bm{\Lambda}'(x)=\textrm{exp}[ig\bm{\omega}(x)]\bm{\Lambda}(x)=\textrm{exp}[igT_a\omega^a(x)]\bm{\Lambda}(x)
\end{equation}
where the $T_a$ satisfy the Lie algebraic commutation relations corresponding to $SU(N)$ and form the natural representation of the generators of the group. The stated Lagrangian, Eqn.\ (\ref{eqn:fourpointoneinelkosolopaper}), is not invariant under such a gauge transformation so we seek to replace the partial derivatives with covariant derivatives of the form
\begin{equation}
D^{\mu}=\partial^{\mu}+ig\mathbf{G}^{\mu}(x)=\partial^{\mu}+igT_aG^{a\mu}(x),
\end{equation}
such that
\begin{equation}
D^{\mu'}=\partial^{\mu}+ig\mathbf{G}^{\mu'}(x)=\partial^{\mu}+ig\mathbf{G}^{\mu}(x)+ig\delta\mathbf{G}^{\mu}(x).
\end{equation}
The Lagrangian
\begin{equation}
\mathcal{L}=D_{\mu}\bm{\stackrel{\neg}\Lambda}D^{\mu}\bm{\Lambda}-m^2\bm{\stackrel{\neg}\Lambda}\bm{\Lambda}
\end{equation}
will be gauge invariant under $SU(N)$ if there can be found a $\delta\mathbf{G}^{\mu}(x)$ such that 
\begin{equation}
\label{eqn:deltaggaugeblahblah}
D^{\mu'}\bm{\Lambda}'(x)=\textrm{exp}[ig\bm{\omega}(x)]D^{\mu}\bm{\Lambda}(x)
\end{equation}
together with
\begin{equation}
\label{eqn:nonabeliangaugeconditionone}
\stackrel{\neg}{\bm{\Lambda}}'(x)=\stackrel{\neg}{\bm{\Lambda}}(x)\textrm{exp}[-ig\bm{\omega}(x)]
\end{equation}
and
\begin{equation}
\label{eqn:nonabeliangaugeconditiontwo}
\stackrel{\neg}{D_{\mu}'\bm{\Lambda}'}(x)=\stackrel{\neg}{D_{\mu}\bm{\Lambda}}(x)\textrm{exp}[-ig\bm{\omega}(x)].
\end{equation}
We emphasize that the gauge invariance of non-abelian Elko Lagrangians relies on Eqn.\ (\ref{eqn:nonabeliangaugeconditionone}) and Eqn.\ (\ref{eqn:nonabeliangaugeconditiontwo}) holding. We believe this should hold, as in our view the way we plug the holes in the incomplete Elko Field Theory is the most natural.

By considering the infinitesimal case, explicit expansion and comparison of the left hand side with the right hand side of this gauge condition Eqn.\ (\ref{eqn:deltaggaugeblahblah}) reveals that it is satisfied if
\begin{equation}
\label{eqn:generalsuntransformationlaw}
\delta\mathbf{G}^{\mu}(x)=-(\partial^{\mu}\bm{\omega}(x))-ig[\mathbf{G}^{\mu}(x),\bm{\omega}(x)].
\end{equation}
The field strength tensor $\mathbf{G}_{\mu\nu}$ follows from the definition
\begin{equation}
[D_{\mu},D_{\nu}]\equiv ig\mathbf{G}_{\mu\nu}.
\end{equation}
Explicit calculation yields the field strength tensor to be
\begin{equation}
\mathbf{G}_{\mu\nu}=T_aG^a_{\mu\nu}=\partial_{\mu}\mathbf{G}_{\nu}-\partial_{\nu}\mathbf{G}_{\mu}-g[\mathbf{G}_{\mu},\mathbf{G}_{\nu}].
\end{equation}
The field strength tensor transforms as
\begin{equation}
\delta G_{\mu\nu}^a=[\bm{\omega},\mathbf{G}_{\mu\nu}]^a.
\end{equation}
We may construct a Lagrangian $\mathcal{L}_G(\mathbf{G},\partial_{\mu}\mathbf{G})$ by varying the Lagrangian and demanding that $\delta\mathcal{L}_G=0$. The form of the solution is standard, and is
\begin{equation}
\mathcal{L}_G=\frac{-1}{4}\mathbf{G}_{\mu\nu}\cdot\mathbf{G}^{\mu\nu}.
\end{equation}
Thus, the form of the Elko Lagrangian $\mathcal{L}(\bm{\Lambda},D_{\mu}\bm{\Lambda},\bm{\stackrel{\neg}\Lambda},D_{\mu}\bm{\stackrel{\neg}\Lambda})$ involving interactions with (massless) $SU(N)$ gauge fields $\mathbf{G}^{\mu}(x)$ is
\begin{equation}
\mathcal{L}=\partial_{\mu}\bm{\stackrel{\neg}\Lambda}\partial^{\mu}\bm{\Lambda}-m^2\bm{\stackrel{\neg}\Lambda}\bm{\Lambda}-j^{\mu}_{a}G_{\mu}^a-\frac{1}{4}\mathbf{G}_{\mu\nu}\cdot G^{\mu\nu}.
\end{equation}
The symmetry current $j^{\mu}_a$ is given by
\begin{equation}
j^{\mu}_a=\frac{-\partial\mathcal{L}}{\partial G_{\mu}^a}=ig[\bm{\stackrel{\neg}\Lambda}T_a(\partial^{\mu}\bm{\Lambda})-(\partial^{\mu}\bm{\stackrel{\neg}\Lambda})T_a\bm{\Lambda}].
\end{equation}
We now describe the specific case of $SU(2)\times U(1)$ local gauge invariance of the Elko Lagrangian. We then discuss Elko electroweak interactions with standard model matter. 

Here we take Elko doublets $\bm{\Lambda}^T=(\Lambda_1,\Lambda_2)$ and form the Lagrangian
\begin{equation}
\mathcal{L}=D_{\mu}\bm{\stackrel{\neg}\Lambda}D^{\mu}\bm{\Lambda}-m^2\bm{\stackrel{\neg}\Lambda}\bm{\Lambda}
\end{equation}
where now the covariant derivative $D^{\mu}$ is defined by 
\begin{equation}
D^{\mu}=\partial^{\mu}+ig\mathbf{W}^{\mu}(x)+ig'B^{\mu}(x).
\end{equation}
The $\mathbf{W}^{\mu}(x)=T_aW^{a\mu}(x)$ are $SU(2)$ gauge fields with $T_a$ being the Lie algebra generators in the natural representation, and $B^{\mu}(x)$ are $U(1)$ abelian gauge fields, the Lagrangian above is invariant under $SU(2)\times U(1)$ local gauge transformations of the form
\begin{equation}
\bm{\Lambda}(x)\rightarrow\bm{\Lambda}'(x)=\textrm{exp}[ig\bm{\eta}(x)+ig'\chi(x)]\bm{\Lambda}(x)
\end{equation}
if\footnote{We here follow the general $SU(N)$ case leading to Eqn.\ (\ref{eqn:generalsuntransformationlaw}).} the transformed gauge fields take the form
\begin{equation}
\mathbf{W}^{\mu}(x)\rightarrow\mathbf{W}^{\mu'}(x)=\mathbf{W}^{\mu}(x)-\partial^{\mu}\bm{\eta}-ig[\mathbf{W}^{\mu},\bm{\eta}]
\end{equation}
\begin{equation}
B^{\mu}(x)\rightarrow B^{\mu'}(x)=B^{\mu}(x)-\partial^{\mu}\chi.
\end{equation}
With these transformation laws, the gauge condition
\begin{equation}
D^{\mu'}\bm{\Lambda}'(x)=\textrm{exp}[ig\bm{\eta}(x)+ig'\chi(x)]D^{\mu}\bm{\Lambda}(x)
\end{equation}
is satisfied. For the $U(1)$ piece, we can define the $U(1)$ field strength tensor $F_{\mu\nu}$ by
\begin{equation}
F_{\mu\nu}=\frac{-i}{g'}[\partial_{\mu}+ig'B_{\mu},\partial_{\nu}+ig'B_{\nu}],
\end{equation}
and also define the $SU(2)$ field strength tensor $\mathbf{W}_{\mu\nu}$ to be
\begin{equation}
\mathbf{W}_{\mu\nu}=\frac{-i}{g}[\partial_{\mu}+ig\mathbf{W}_{\mu}(x),\partial_{\nu}+ig\mathbf{W}_{\nu}(x)].
\end{equation}
We can now form the Lagrangian
\begin{equation}
\mathcal{L}=\partial_{\mu}\bm{\stackrel{\neg}\Lambda}\partial^{\mu}\bm{\Lambda}-m^2\bm{\stackrel{\neg}\Lambda}\bm{\Lambda}-g'j_B^{\mu}B_{\mu}-gj^{\mu}_{\omega,a}W_{\mu}^a-\frac{1}{4}F_{\mu\nu}F^{
mu\nu}-\frac{1}{4}\mathbf{W}_{\mu\nu}\cdot\mathbf{W}^{\mu\nu}
\end{equation}
where the symmetry currents $j_B^{\mu}$ and $j^{\mu}_{\omega,a}$ are defined by
\begin{equation}
\frac{-\partial\mathcal{L}(\bm{\Lambda},\bm{\stackrel{\neg}\Lambda},D_{\mu}\bm{\Lambda},\bm{\stackrel{\neg}\Lambda})}{\partial B_{\mu}}\quad\textrm{and}\quad\frac{-\partial\mathcal{L}(\bm{\Lambda},\bm{\stackrel{\neg}\Lambda},D_{\mu}\bm{\Lambda},\bm{\stackrel{\neg}\Lambda})}{\partial \mathbf{W}_{\mu}}
\end{equation}
respectively. 

As formulated here, the gauge fields $B_{\mu}$ and $\mathbf{W}_{\mu}$ are massless since any mass term is not invariant under gauge transformations
\begin{equation}
m^2B_{\mu}'B^{\mu'}\neq m^2B_{\mu}B^{\mu}
\end{equation}
 and similarly for the $SU(2)$ gauge field mass terms. Just like with the standard model case, these gauge fields can be given mass by interaction with the Higgs field. The standard Higgs mechanism is described, in, for example, [\onlinecite[sec.14.4]{aitchison1989gauge}][\onlinecite[ch.8]{ryder1985quantum}][\onlinecite[ch.21]{weinberg1996quantum2}], where the (unphysical) fields $W^{3\mu}$ and $B^{\mu}$ are transformed into two new (physical) fields $Z^{\mu}$ and $A^{\mu}$ via the transformations
\begin{equation}
W^{3\mu}=\cos(\theta_W)Z^{\mu}+\sin(\theta_W)A^{\mu}\quad\textrm{and}\quad B^{\mu}=-\sin(\theta_W)Z^{\mu}+\cos(\theta_W)A^{\mu}
\end{equation}
involving the Weinberg angle $\theta_W$. These transformations together with the Higgs doublet
\begin{equation}
\phi=\frac{1}{\sqrt{2}}\left(\begin{array}{cccc}
0\\
f+H(x)
\end{array}\right)
\end{equation}
with $f$ a number and $H$ the physical Higgs field, have the effect of giving mass to the $W$ and $Z$ fields.
\section{Discussion of Elko Gauage Invariance and Elko Field Theory Incompleteness}
Our results are consistent with the claim that Elko fields can couple to the Higgs field, and the claim that Elko fields can have quartic self-interactions. They do however contradict the assertion made in [\onlinecite[p.7]{ahluwalia2008local}] that \textit{$\mathcal{L}^{\Lambda}(x)$ and $\mathcal{L}^{\lambda}(x)$ do not carry invariance under Standard Model gauge transformations.} The justification for this claim made in the cited paper was made from considerations at the \textit{spinor} level, considering Elko spinors as eigenspinors of an antilinear $(\frac{1}{2},0)\oplus(0,\frac{1}{2})$ representation space operator $S(C)$. We take the view that at the level of the \textit{state space}, the spinor coefficients present in the quantum field operator do not determine whether a Lagrangian composed of \textit{quantum field operators} is invariant under a gauge transformation. An argument of the nature described in [\onlinecite[p.7]{ahluwalia2008local}] applies if we consider Lagrangians composed of Elko \textit{spinors} where the Lagrangian is then defined in a classical spinor space and is not viewed as an operator in Hilbert space. We believe that this result does not automatically nullify the gauge invariance of Lagrangians that are operators on the Hilbert space of physical states, composed of Elko \textit{quantum field operators}.

As pointed out in Sec.\ (\ref{sec:elkoreviewanddarkness}), another point raised in [\onlinecite{ahluwalia2008local}] was that Elko fields must be dark with respect to (non-Higgs) Standard Model particles on account of the Elko mass dimensionality being different from the mass dimensionality of the Standard Model fermions, thus forbidding Elko fields from being allowed to enter into the fermionic doublets of the Standard Model. We agree that Elko fields cannot enter the Standard Model doublets. However we do \textit{not} agree that this implies that Elko fields are dark with respect to these Standard Model doublets. The reason for our divergent viewpoint is as follows:

As we have already pointed out in this section, it is possible at the level of the Hilbert space of physical states for Lagrangian operators composed of Elko quantum field operators to be gauge invariant. Once we see this, we can then take Elko fields through the gauge process and hence it is possible to set up Elko doublets and Elko Lagrangians that are invariant under $SU(2)\times U(1)$ gauge transformations. There is no reason why we cannot also choose to take the left-handed Elko components to form the doublets. We can then (even without demanding that we form the Elko doublets purely from the left-handed components of Elko fields) automatically write down symmetry currents in the usual way that arise from the invariance of the Lagrangian under $SU(2)\times U(1)$ gauge transformations. Once we have written down an Elko symmetry current consisting of doublets whose entries are solely Elko fields, we can couple this symmetry current four vector operator with the Standard Model symmetry current four vector operators and form \textit{new} interaction terms in an expanded Hamiltonian density operator that now admits new additional interactions; specifically, interactions between the Elko fields and the Standard Model fields. This is at the level of the Electroweak Theory, involving the $W^{\pm}$ and $Z^0$ vector bosons. By similar reasoning, simpler arguments apply in the simpler theory of Quantum Electrodynamics. There are also no clear reasons why we could not also consider Elko triplets transforming under the natural representation of the $SU(3)$ gauge group and couple the corresponding Elko symmetry currents to the symmetry currents in Quantum Chromodynamics.

At this point, having made the case for Elko in principle being able to interact with the usual Standard Model particles (not just the Higgs particle) we nonetheless claim that Elko is a viable dark matter candidate. We say this for three (albeit to an extent interelated) reasons.

The first reason that Elko particles might still be viable dark matter candidates is that in the particular sense described in Sec.\ (\ref{sec.elkopossiblenoncausality}), Elko fields have an element of  non-locality, even along the axis of locality. If a particle is not local, it is not clear to us exactly what this implies when it comes to the issue of how we can detect such a particle. The non-local nature of Elko fields may give Elko particles the \textit{appearance} of being dark in the sense that we have trouble finding them, even though they admit gauge interactions.

The second reason that Elko particles might still be viable dark matter candidates is that Elko fields are not quantum fields in the sense of Weinberg (see [\onlinecite{Conferenceproceedings}]), and that, more specifically, they do not transform correctly under the Lorentz group. They break rotational invariance in particular [\onlinecite{Conferenceproceedings,ahluwalia2010elko,gillardmartinpaperv2}]. This implies that if we were to construct Hamiltonian densities out of the Elko fields $\Lambda_{\ell}(x)$ and their complex conjugate transpose adjoint's $\Lambda^{\dagger}_{\ell}(x)$, we would not have a Hamiltonian density which is a Lorentz scalar under rotations. If we were then to couple a Standard Model symmetry current with an Elko symmetry current, the resulting object may \textit{look} like a scalar when in reality it is not a scalar under rotations. This in turn may affect the detectability of Elko particles.

In this paper therefore, we take what we see as the most natural positions, given the stated ambiguities and holes in Elko Field Theory, and  give the natural consequences of plugging the holes in Elko Field Theory in what we see as the most natural way.
\section{Elko's Electroweak Interactions with Standard Model Matter}
Here we examine some basic consequences of Elko particles carrying weak isospin. First we consider the left-handed aspect of the weak interactions. For standard Dirac-type fermions, the wave equation for a free Dirac field $\psi$ is
\begin{equation}
(i\gamma^{\mu}\partial_{\mu}-m)\psi=0.
\end{equation}
If we define the projection operators $P_L=\frac{1}{2}(1-\gamma_5)$ and $P_R=\frac{1}{2}(1+\gamma_5)$ and apply $P_L$ to the left side of the Dirac equation, and note that $\{\gamma^{\mu},\gamma_5\}=0$, we see that
\begin{equation}
i\gamma^{\mu}\partial_{\mu}\psi_R=m\psi_L
\end{equation} 
where $P_R\psi\equiv\psi_R$ and $P_L\psi\equiv\psi_L$. Similarly, if we apply the projection operator $P_R$ to the Dirac equation, we get
\begin{equation}
i\gamma^{\mu}\partial_{\mu}\psi_L=m\psi_R.
\end{equation}
Given that 
\begin{equation}
m\psi=m\psi_L+m\psi_R,
\end{equation}
it follows that the Dirac equation cannot be locally invariant under the \textit{left-handed} gauge group $SU(2)_L$. The standard solution to this problem for standard Dirac fermions is to say that free fermions are actually massless and that these massless fermion fields acquire mass by interacting with the Higgs field via the \textit{Yukawa} interactions (see, for example, [\onlinecite[p.465]{aitchison1989gauge}]):
\begin{equation}
i\gamma^{\mu}\partial_{\mu}\psi_R=g_f\phi^{\dagger}l,\quad i\gamma^{\mu}\partial_{\mu}l=g_f\phi\psi_R,
\end{equation}
where $l$ is a doublet with non-zero isospin and $g_f$ is a coupling constant.
In this way, the Higgs boson takes on an additional very important role in the Standard Model. In addition to giving mass to the weak force gauge quanta, the Higgs field also gives mass to all of the Dirac type fermions of the Standard Model.

If we make the assumption that the weak force interacts only with left-handed fermions regardless of whether they are Standard Model fermions, or non-Standard Model fermions like Elko, then we need to examine Elko fields to see whether left and right-handed components get mixed like they were in the Dirac field. The dynamical equation of motion for the Elko field is Klein-Gordon in form:
\begin{equation}
(\partial_{\mu}\partial^{\mu}+m^2)\Lambda(x)=0.
\end{equation}
Applying the projection operators to this equation gives
\begin{equation}
(\partial_{\mu}\partial^{\mu}+m^2)\Lambda_L=0\quad\textrm{and}\quad (\partial_{\mu}\partial^{\mu}+m^2)\Lambda_R=0
\end{equation}
so we see that Elko left and right components are \textit{not} mixed up as they are in the Dirac case. We therefore come to a simple but profound difference between \textit{Elko}-type fermionic fields and standard \textit{Dirac}-type fermionic fields. Elko free particle states do not need to acquire mass from interacting with the Higgs boson, in contrast to Dirac free particle states, which are thought to be intrinsically massless and need to interact with the Higgs field in order to acquire what is normally regarded as \textit{their} rest mass. If Elko particles were to acquire their mass from the Higgs field also, the form of the interaction would be fundamentally different.

Also, Elko left and right-handed components can be separated leaving the left-handed components to transform differently under $SU(2)_L\times U(1)$ from the right-handed components. Thus, Elko passes a key test concerning its likelihood of being able to interact electroweakly with Standard Model matter. The right-handed components will transform as \textit{singlets} under the electroweak gauge group where as the left-handed components will transform as \textit{doublets} under the electroweak gauge group. We now explore left-handed Elko doublets having non-zero isospin.

In order for an Elko doublet to interact with a $W^+$ or $W^-$ particle, we require that there be an electric charge difference of unity between the top and bottom components of the weak isospinor. The Elko Lagrangian accommodates global $U(1)$ gauge symmetries so Elko particles can, in principle, carry electric charge. Hence we may write down an Elko doublet $\mathcal{E}$ of the form
\begin{equation}
\mathcal{E}=\left(\begin{array}{cccc}
\Lambda^0\\
\Lambda^-
\end{array}\right)_L
\end{equation}
where the superscript ``0'' denotes an electrically neutral Elko field and the superscript ``-'' denotes an electrically charged Elko quantum field differing from the neutral Elko field by one unit of electric charge.

A possible Elko-Standard Model electroweak interaction might therefore look something like
\begin{equation}
\nonumber \Lambda^-\rightarrow \Lambda^0+W^-+X
\end{equation}
where $X$ denotes other decay products adding no net charge to the interaction that might not necessarily consist of purely Standard Model particles. If it turns out that Elko, on account of violating rotational symmetries with its axis of locality, is dark with respect to usual direct detection methods, a possible decay like the one above may give us a possible way of inferring Elko's presence \textit{indirectly} by studying the Standard Model component of the interactions that we can detect directly.

The particular example of a possible interaction may very well come across as the \textit{appearance} of the violation of conservation of electric charge if the special non-locality attributes of Elko fields complicate how and where the electric fields are produced. If electric charge conservation ever did appear to be broken by particle interactions involving $W$ vector bosons, we would have evidence, not necessarily that electric charge conservation has failed to be a universal law of physics, but that Elko particles might be making their presence felt.

At presently accessible energies it may turn out that such an interaction as that stated above might not ever be observed. This would suggest that either Elko particles do not exist, or, that there is some as yet unexplored physical mechanism (perhaps some sort of spontaneously broken symmetry mechanism) which prevents Elko symmetry currents from coupling to the charged electroweak symmetry currents. If Elko particles do constitute at least a large part of the dark matter sector, then there is evidence [\onlinecite{ahluwalia2010very}] that Elko should already be comfortably within the energy range accessible to experiments. Another possibility is that there is some mechanism that restricts the Elko electroweak symmetry currents to couple only with the neutral electroweak symmetry currents involving the $Z^0$ vector boson.

The first experimental evidence for the $Z^0$ particle was found using the scattering reaction $\bar{\nu}_{\mu}+e\rightarrow\bar{\nu}_{\mu}+e$ [\onlinecite{arnison1983experimental}]. The $Z^0$ particle can interact with any Standard Model particles except for gluons and photons. We take the view that since Elko particles could, in principle, carry non-zero isospin, they should also be able to interact electroweakly by coupling to the neutral currents. Elko particles should be able to scatter off each other and exchange a $Z^0$ particle.

Explicitly, the Elko current takes the form
\begin{equation}
j^{\mu}_a=\frac{ig}{2}[\bm{\stackrel{\neg}\Lambda}\tau_a(\partial^{\mu}\bm{\Lambda})-(\partial^{\mu}\bm{\stackrel{\neg}\Lambda})\tau_a\bm{\Lambda}].
\end{equation}
where the $\frac{\tau_a}{2}$ are the generators of the weak isospin group $SU(2)_L$ in the natural representation, numerically identical to the Pauli spin matrices $\bm{\sigma}$.
\section{Elements of Elko Non-Locality} \label{sec.elkopossiblenoncausality}
In the Standard Model, to every matter field operator $\psi_{\ell}(x)$ acting on the Hilbert space of physical states, there corresponds a unique field operator, the adjoint $\psi_{\ell}^{\dagger}(x)$, which is also a matter field operator, with clear physical interpretation. In order for both matter field operators to be \textit{causal}, they have to anticommute (or commute in the case of bosons) with themselves and each other at spacelike separated $x$ and $x'$ [\onlinecite[p.198]{Weinberg1996quantum1}]:
\begin{equation}
\{\psi_{\ell}(x),\psi_{\ell'}(x')\}=\{\psi^{\dagger}_{\ell}(x),\psi^{\dagger}_{\ell'}(x')\}=\{\psi_{\ell}(x),\psi^{\dagger}_{\ell'}(x')\}=0.
\end{equation}
It is clear from the Elko literature (see [\onlinecite[p.6]{ahluwalia2008local}] for example) that the Elko fields anticommute with themselves at spacelike separation but not clear as to whether they anticommute with their \textit{adjoints} at spacelike separation. In this section we take a look at the anticommutator of the Elko field with its adjoint at spacelike separation. Since we think of the Elko field operator as an operator on Hilbert space that destroys Elko particles and create antiparticles, it follows that its unique adjoint has the physical interpretation of creating Elko particles and destroying Elko antiparticles.

Here, we take a general Elko field with four rest spinors of the form:
\begin{eqnarray}
\xi\left(\mathbf{0},\frac{1}{2}\right)=\left(\begin{array}{cccc}
-\eta b_1^*\\
\eta a_1^*\\
a_1\\
b_1
\end{array}\right),\qquad\xi\left(\mathbf{0},\frac{-1}{2}\right)=\left(\begin{array}{cccc}
-\eta b_2^*\\
\eta a_2^*\\
a_2\\
b_2
\end{array}\right)
\end{eqnarray}
\begin{eqnarray}
\zeta\left(\mathbf{0},\frac{1}{2}\right)=\left(\begin{array}{cccc}
-\eta d_1^*\\
\eta c_1^*\\
c_1\\
d_1
\end{array}\right),\qquad\zeta\left(\mathbf{0},\frac{-1}{2}\right)=\left(\begin{array}{cccc}
-\eta d_2^*\\
\eta c_2^*\\
c_2\\
d_2
\end{array}\right)
\end{eqnarray}
for general complex numbers $a_1$, $a_2$, $b_1$, $b_2$, $c_1$, $c_2$, $d_1$, $d_2$ and phase $\eta$. We look at the anticommutator of the field $\Lambda_{\ell}(x)$ at spacetime point $x$ with its adjoint $\Lambda_{\bar{\ell}}^{\dagger}(y)$ at spacetime point $y$ which is at spacelike separation from the point $x$. Elko rest spinors are of this form except with the additional restriction that the top two components and bottom two components of each Elko spinor are eigenstates of the two-dimensional spin-$\frac{1}{2}$ helicity operator $\frac{1}{2}\bm{\sigma}\cdot\hat{\mathbf{p}}$. We here place no such restrictions on these constants, and show that even this more general class of spinors cannot serve as suitable sets of coefficient functions $\xi_{\ell}(\mathbf{p},\alpha)$ and $\zeta_{\ell}(\mathbf{p},\alpha)$ to produce a field operator satisfying the stated causality condition.

A calculation of the anticommutator between the field operator $\Lambda_{\ell}(x)$ at $x$ and its adjoint $\Lambda_{\bar{\ell}}^{\dagger}(y)$ at $y$ yields the general form:
\begin{equation}
\{\Lambda_{\ell}(x),\Lambda_{\bar{\ell}}^{\dagger}(y)\}=\int\frac{d^3\mathbf{p}}{(2\pi)^3}\sum_{\alpha}\left[\xi_{\ell}(\mathbf{p},\alpha)\xi_{\bar{\ell}}^{\dagger}(\mathbf{p},\alpha)e^{ip\cdot(x-y)}+\zeta_{\ell}(\mathbf{p},\alpha)\zeta_{\bar{\ell}}^{\dagger}(\mathbf{p},\alpha)e^{-ip\cdot(x-y)}\right].
\end{equation}
In order for these anticommutator components to vanish, they must do so for all combinations of the pair $(\ell,\bar{\ell})$. 

The spin sums are quite messy but it is sufficient for our purposes just to examine the $(\ell,\bar{\ell})=(1,1)$ component and look at $\{\Lambda_1(x),\Lambda_1^{\dagger}(y)\}$. If we calculate the spin sums $\left[\sum_{\alpha}\xi(\mathbf{p},\alpha)\xi^{\dagger}(\mathbf{p},\alpha)\right]_{11}$ and $\left[\sum_{\alpha}\zeta(\mathbf{p},\alpha)\zeta^{\dagger}(\mathbf{p},\alpha)\right]_{11}$ and substitute them into the expression for $\{\Lambda_1(x),\Lambda_1^{\dagger}(y)\}$ and make the standard substitution $p_{\mu}\rightarrow-i\partial_{\mu}$ we get
\begin{equation}
\label{eqn:bigmess}
[-(m-i\partial_0-i\partial_z)(-i\partial_x+i(-i\partial_y))(b_1^*a_1+b_2^*a_2)-(m-i\partial_0-i\partial_z)(-i\partial_x-i(-i\partial_y))\times
\end{equation}
\begin{equation}
\nonumber(a_1^*b_1+a_2^*b_2)+(m-i\partial_0-i\partial_z)(m-i\partial_0-i\partial_z)(b_1^*b_1+b_2^*b_2)-(\partial_x^2+\partial_y^2)(a_1^*a_1+a_2^*a_2)]H(x-y)+
\end{equation}
\begin{equation}
\nonumber[-(m-i\partial_0-i\partial_z)(-i\partial_x+i(-i\partial_y))(d_1^*c_1+d_2^*c_2)-(m-i\partial_0-i\partial_z)(-i\partial_x-i(-i\partial_y))\times
\end{equation}
\begin{equation}
\nonumber(c_1^*d_1+c_2^*d_2)+(m-i\partial_0-i\partial_z)(m-i\partial_0-i\partial_z)(d_1^*d_1+d_2^*d_2)-(\partial_x^2+\partial_y^2)(c_1^*c_1+c_2^*c_2)]H(y-x).
\end{equation}

where $H(x-y)$ is an even function of $x-y$ for spacelike intervals:
\begin{equation}
H(x-y)=\int\frac{d^3\mathbf{p}}{(2\pi)^32p_0(m+p_0)}e^{-ip\cdot (x-y)}.
\end{equation}
By inspection of Eqn.\ (\ref{eqn:bigmess}) we deduce that
\begin{equation}
a_1^*a_1+a_2^*a_2=-c_1^*c_1-c_2^*c_2
\end{equation}
and
\begin{equation}
b_1^*b_1+b_2^*b_2=-d_1^*d_1-d_2^*d_2.
\end{equation}
The only solution to this however, is the trivial solution:
\begin{equation}
a_1=a_2=b_1=b_2=c_1=c_2=d_1=d_2=0.
\end{equation}
so in general
\begin{equation}
\{\Lambda_{\ell}(x),\Lambda_{\bar{\ell}}^{\dagger}(y)\}\neq0
\end{equation}
for a spacelike interval $(x-y)^2<0$.

Elko fields thus have an element of acausality to them \textit{even along} the axis of locality. This make is less clear how Elko particles and the Standard Model particles they interact with, manifest themselves during Elko particle interactions. An analysis is urgently needed to determine exactly what physical effects these Elko axes of locality and possible Elko non-causality imply, for Elko detectability and Elko interactivity with other particles, especially the Standard Model particles.
\section{Conclusion}
We have shown that Elko Lagrangians are gauge invariant under what we consider to be the most natural way of completing the Elko Field Theory, which we explained was incomplete. We have shown that in principle, Elko fields could interact electroweakly with Standard Model matter and that the Elko $SU(2)\times U(1)$, symmetry currents can couple to the symmetry currents associated with the Standard Model doublets. Elko's violation of rotational symmetries and, its element of non-locality, even along the axis of locality, may still keep Elko particles appearing dark with respect to direct detection, thus accounting for why Elko has evaded detection even though it is free to interact with Standard Model matter. Indirect detection of Elko particles by studying Standard Model matter components of Elko field interactions could provide the best way of most easily detecting Elko particles.

There are many possibilities that need to be explored. The usual Dirac spinors have their quantum field counterparts. The Elko spinors have proved potentially very useful in a variety of areas of physics research, as pointed out at length in our introduction to this paper. This provides good motivation to more fully develop the associated Elko \textit{Quantum Field Theory} to the point where specific experiments can be done. Now that the Large Hadron Collider is available to probe higher energies than were previously accessible, this would be an ideal time to put Elko \textit{fields} on a more complete theoretical footing. Much remains to be done. Cross sections and branching ratios need to be calculated, and the crucial renormalizability aspect of the Elko Quantum Field Theory needs to be studied.

The issue of the incompleteness of Elko Field Theory has far reaching consequences beyond whether Elko fields can be derived in the non-standard Wigner classes. The incompleteness of the theory regarding the Elko dual operation on operators in Hilbert space greatly obscures whether Elko fields can admit gauge interactions. If Eqn.\ (\ref{eqn:criticalforelkogaugeinvariance}) holds, then there is no reason why Elko fields cannot interact with Standard Model gauge quanta, and therefore participate in quantum electrodynamic, quantum chromodynamic and electroweak interactions. The \textit{nature} of the incompleteness of Elko Field Theory therefore constitutes a potentially significant criticism of the theory. The ability for Elko particles, interpreted in what we consider to be the most natural way to admit such a wide array of interactions with Standard Model particles, reduces their suitability as dark matter candidates, but does not rule Elko particles out as dark matter candidates for several reasons. Firstly, since Elko fields have a single axis of locality (at most), they could be spread out over vast distances, creating all sorts of poorly understood complications when looking for Elko particles in a small area, such as the LHC for example. Secondly, the lack of observations of Elko interacting with the Standard Model gauge quanta could be interpreted as evidence that the holes in Elko Field Theory should be filled in less obvious ways. Specifically, one might decide that we do not want Eqn.\ (\ref{eqn:criticalforelkogaugeinvariance}) to hold, not for any theoretical reasons, but simply because of the lack of Elko particle observations in nature.




%




%











%







\begin{acknowledgments}

 I thank the University of Canterbury Physics and Astronomy Department for financial support.

\end{acknowledgments}



%

\end{document}